\documentclass[journal]{IEEEtran} %letter
\IEEEoverridecommandlockouts

\usepackage[colorlinks]{hyperref}
\usepackage{mathrsfs}
\usepackage{graphicx,color,overpic,psfrag}
\usepackage{amsmath, amssymb}
\usepackage{latexsym}
\usepackage{bm}
\usepackage{amssymb}
\usepackage{cases}
\usepackage{array}
\usepackage{fancyhdr}
\usepackage{setspace}
\usepackage{caption}
\usepackage{indentfirst}
\usepackage{cases}
\usepackage{url}
\usepackage{algpseudocode}
\usepackage{algorithm}
\usepackage{blkarray}
\usepackage{subfig}
\usepackage{graphicx}

\usepackage{booktabs}
\usepackage{multirow}
\usepackage{dsfont}
\usepackage{tabularx}
\usepackage[table]{xcolor}
\usepackage{amsfonts}
\usepackage{amsthm}
\usepackage{stfloats}
\usepackage{letltxmacro}
\usepackage{lettrine}

\newcommand{\ba}{\mathbf{a}}
\newcommand{\bb}{\mathbf{b}}
\newcommand{\bc}{\mathbf{c}}

\newcommand{\bn}{\mathbf{n}}

\newcommand{\br}{\mathbf{r}}
\newcommand{\bs}{\mathbf{s}}

\newcommand{\bv}{\mathbf{v}}
\newcommand{\bw}{\mathbf{w}}
\newcommand{\bx}{\mathbf{x}}
\newcommand{\by}{\mathbf{y}}

\newcommand{\bA}{\mathbf{A}}

\newcommand{\bE}{\mathbf{E}}

\newcommand{\bI}{\mathbf{I}}
\newcommand{\bJ}{\mathbf{J}}

\newcommand{\bR}{\mathbf{R}}

\newcommand{\bW}{\mathbf{W}}

\graphicspath{{figure/}}
\allowdisplaybreaks

\begin{document}
	
	\title{\LARGE SCNR Maximization for MIMO ISAC Assisted by Fluid Antenna System}

	\author{ Yuqi~Ye, 
            Li~You,~\IEEEmembership{Senior~Member,~IEEE,} 
            Hao Xu,~\IEEEmembership{Senior~Member,~IEEE,} 
            Ahmed Elzanaty,~\IEEEmembership{Senior~Member,~IEEE,} 
            Kai-Kit Wong,~\IEEEmembership{Fellow,~IEEE,} 
            and Xiqi~Gao,~\IEEEmembership{Fellow,~IEEE}%~\IEEEmembership{Fellow,~IEEE}% <-this % stops a space
\vspace{-9mm}			
		
		\thanks{
%This work was supported by the National Key Research and Development
%			Program of China under Grant 2018YFB1801103, the Key Technologies
%			R\&D Program of Jiangsu (Prospective and Key Technologies for Industry)
%			under Grants BE2022067 and BE2022067-5, the Jiangsu Province Basic
%			Research Project under Grant BK20192002, the Fundamental Research Funds
%			for the Central Universities under Grant 2242021R41148, and the Young Elite
%			Scientist Sponsorship Program by China Institute of Communications. The associate editor coordinating the review of this paper and approving it for publication was Dr. Yingyang Chen. \emph{(Corresponding author: Li You.)}

Yuqi Ye, Li You, and Xiqi Gao are with the National Mobile Communications Research Laboratory, Southeast University, Nanjing 210096, China, and also with the Purple Mountain Laboratories, Nanjing 211100, China (e-mail: $\rm yqye@seu.edu.cn$; $\rm lyou@seu.edu.cn$;  $\rm xqgao@seu.edu.cn$).

Hao Xu is with the National Mobile Communications Research Laboratory, Southeast University, Nanjing 210096, China (e-mail: $\rm hao.xu@seu.edu.cn$).

Kai-Kit Wong is affiliated with the Department of Electronic and Electrical Engineering, University College London, Torrington Place, WC1E 7JE, United Kingdom and he is also affiliated with Yonsei Frontier Lab, Yonsei University, Seoul, Korea (e-mail: $\rm kai\text{-}kit.wong@ucl.ac.uk$).

Ahmed Elzanaty is with the 5GIC and 6GIC, Institute for Communication Systems, University of Surrey, GU2 7XH Guildford, U.K (e-mail:
$\rm a.elzanaty@surrey.ac.uk$).}
}

\maketitle
\begin{abstract}
The integrated sensing and communication (ISAC) technology has been extensively researched to enhance communication rates and radar sensing capabilities. Additionally, a new technology known as fluid antenna system (FAS) has recently been proposed to obtain higher communication rates for future wireless networks by dynamically altering the antenna position to obtain a more favorable channel condition. The application of the FAS technology in ISAC scenarios holds significant research potential. In this paper, we investigate a FAS-assisted multiple-input multiple-output (MIMO) ISAC system for maximizing the radar sensing signal-clutter-noise ratio (SCNR) under communication signal-to-interference-plus-noise ratio (SINR) and antenna position constraints. We devise an iterative algorithm that tackles the optimization problem by maximizing a lower bound of SCNR with respect to the transmit precoding matrix and the antenna position. By addressing the non-convexity of the problem through this iterative approach, our method significantly improves the SCNR. Our simulation results demonstrate that the proposed scheme achieves a higher SCNR compared to the baselines.
		
\end{abstract}

\section{Introduction}
\IEEEPARstart{W}{hile} the fifth-generation (5G) mobile communications has provided unprecedented speeds and connectivity, the sixth-generation (6G) aims to further enhance these capabilities by integrating advanced technologies \cite{wang2023road}. As a key technology for 6G wireless communications, integrated sensing and communication (ISAC) uses the same hardware and spectrum resources for sensing and communication tasks \cite{9852292}. Motivated by the many advantages of ISAC, numerous related research has emerged. In \cite{10217169}, the authors studied the trade-offs between Cram\'{e}r-Rao bound (CRB) and communication rate in multiple-input multiple-output (MIMO) ISAC systems. Earlier, \cite{9724174} investigated generalized transceiver beamforming strategies for dual-functional radar-communication (DFRC) systems. Later, \cite{9913311} tackled the beamforming design for radar-communication systems assisted by reconfigurable intelligent surface (RIS). Nevertheless, traditional fixed-position antennas (FPAs) are limited by the number of available radio-frequency (RF) chains in terms of degree of freedom (DoF) for sensing and communication systems. In communication, FPAs often struggle with adaptability to dynamic environments, leading to suboptimal signal reception and interference mitigation. By contrast, in sensing, FPAs with a fixed aperture lack flexibility to dynamically adjust their configurations, resulting in lower angle resolution and less effective radar target detection.

Fluid antenna system (FAS) presents a promising solution to these challenges and provides a new DoF. Conceptually, it represents shape-flexible, position-flexible antenna technologies for wireless communication \cite{wong2020fluid0,New-submit2024}. FAS comes in the form of fluidic radiating structures \cite{Shen-tap_submit2024}, mechanically-movable antenna structures, reconfigurable RF pixels \cite{Zhang-pFAS2024}, and etc. FAS has garnered considerable attention recently due to its potential to revolutionize wireless communication systems. As a matter of fact, the first analytical results on FAS were presented in \cite{Wong-fas-cl2020} and \cite{wong2020fluid} while \cite{10416896} addressed the antenna position and beamforming design and demonstrated great efficiency gains in FAS. Recently, in \cite{10328751}, the authors investigated the use of fluid antennas in MIMO  systems, leveraging statistical channel state information (CSI) to optimize performance.

For sensing, FAS can provide an adaptive aperture with the same number of RF chains as fixed antennas, leading to higher angle resolution and more accurate radar target detection. For communication, the adaptability of FAS enhances interference mitigation, leading to better overall communication efficiency \cite{wong2022bruce, 10130117}. Despite the recognized importance of both ISAC and FAS in next-generation wireless communications, the potential integration of FAS within ISAC scenarios still needs to be explored. Studying this integration could unveil significant improvements in both communication and sensing capabilities.

Motivated by these considerations, this paper aims to investigate the MIMO ISAC system supported by FAS. The primary focus is on enhancing the radar sensing signal-to-clutter-plus-noise ratio (SCNR) while considering the constraints related to communication signal-to-interference-plus-noise ratio (SINR) and fluid antennas. In particular, we introduce an efficient algorithm designed to optimize both the joint transmit precoding matrix and the fluid antenna positions. Simulation results will show that the proposed FAS-assisted ISAC scheme surpasses the baseline methods in terms of the SCNR performance.

\textit{Notations}: The symbols $\mathbf{A}^H$ and $\mathbf{A}^{-1}$ denote the conjugate transpose and the inverse of the matrix $\mathbf{A}$, respectively. 
For  vector $\mathbf{a}$, $\mathbf{a}^T$ denotes its transpose, while $\mathbf{a}^{*}$ is its conjugate. 
The $l_2$ norm is represented as $\|\cdot\|_2$. Additionally, the notations $\angle\cdot$ and $|\cdot|$ represent the phase and modulus of a complex number, respectively. Notation $\otimes$ denotes the Kronecker product.

\vspace{-2mm}
\section{System Model And Problem Formulation}
Consider a MIMO ISAC system operating in the millimeter wave band, consisting of a base station (BS), $K$ communication users, and one radar target. The BS is equipped with a FPA-based uniform planar array (UPA) of size $M = M_x \times M_y$ for transmit antennas and $N$ fluid antennas for receive antennas. It transmits signals to $K$ single-FPA communication users while also sensing a radar target. Fluid antennas at the BS can switch to any positions within a given $A\times A$ spatial region $\mathcal S$. The coordinates of the $n$-th fluid antenna are written as $\mathbf  r_n=(x_n,y_n)^T$. The positions collection of $N$ fluid antennas is represented as ${\mathbf  r}=[\mathbf  r_1,\dots,\mathbf  r_N]\in\mathbb R^{2\times N}$.

%FAs at the transmitter and receiver are all connected to radio frequency chains via flexible cables so that they can move freely in the given regions $\mathcal {S}_{\rm t}$ and $\mathcal S_{\rm r}$, respectively. Here we introduce the two-dimensional Cartesian coordinate system to describe the specific positions of the transmit and receive FAs. The coordinates of the $n$-th $(1\leq n\leq N)$ transmit FA and $m$-th $(1\leq m\leq M)$ receive FA are denoted as $\mathbf t_n=(x_{n},y_{n})^T\in \mathcal S_{\rm t}$ and $\mathbf  r_m=(x_{m},y_{m})^T\in \mathcal S_{\rm r}$, respectively. We assume $\mathcal S_{\rm t}$ and $\mathcal S_{\rm r}$ are $A\times A$ square regions, i.e., $\mathcal S_{\rm t}=\mathcal S_{\rm r}=[-A/2,A/2]\times[-A/2,A/2]$. The position collections of $N$ transmit FAs and $M$ receive FAs are denoted by ${\mathbf  t}=[\mathbf  t_1,\dots,\mathbf  t_N]\in\mathbb R^{2\times N}$ and ${\mathbf  r}=[\mathbf  r_1,\dots,\mathbf  r_M]\in\mathbb R^{2\times M}$, respectively.

%Note that changing the positions of transmit and receive antennas results in different MIMO channel matrices. Therefore, The assisted channel matrix between the transmitter and receiver can be represented by a function related to the position of the transmit and receive MAs, i.e., $\mathbf H({\mathbf t},{\mathbf r})\in\mathbb C^{M\times N}$. 

\vspace{-2mm}
\subsection{Communication Model}
The transmit signal is denoted by $\mathbf x=\mathbf W \bs\in\mathbb C^{M\times 1}$, in which $\bW=[\bw_1,\dots,\bw_K]\in\mathbb C^{M\times K}$ is the transmit precoding matrix and $\bs$ is the data vector satisfying  $\mathbb{E}\{\bs\bs^H\}=\bI_K$. The signal received by user $k$ is expressed as
\begin{equation}
y_k=\mathbf h_k^H\mathbf W \bs+n_k,
\end{equation}
where $\mathbf h_k\in\mathbb C^{M\times 1}$ is the channel vector between the BS and the $k$-th communication user, and $ n_k\sim\mathcal{CN}(0,\sigma_k^2)$ represents the complex additive white Gaussian noise (AWGN).

Using the Saleh-Valenzuela model \cite{8926431}, the channel vector from the BS to user $k$ is expressed as
\begin{equation}
\mathbf{h}_k=\sqrt{\frac{1 }{L_k}} \sum_{l=1}^{L_k} \rho_{k,l} \mathbf{a}_t\left(\psi_{k,l}, \vartheta_{k,l}\right),
\end{equation}
where $L_k$ is the number of paths between the BS and user $k$, $\psi_{k,l}$ and $\vartheta_{k,l}$ denote the elevation and azimuth angles of departure (AoDs) for the $l$-th path, and $\rho_{k,l}$ represents the corresponding path gain. Since the BS employs a UPA for transmit antennas, the steering vector is given by
\begin{equation}
\begin{aligned}
\mathbf{a}_t(\psi_{k,l},\vartheta_{k,l})=&[1,\dots,e^{\jmath\frac{2\pi}{\lambda} (M_x-1) d\sin\psi_{k,l}\cos\vartheta_{k,l}}]^T\\
&\otimes[1,\dots,e^{\jmath\frac{2\pi}{\lambda}(M_y-1) d\cos\psi_{k,l}}]^T\in \mathbb{C}^{M\times 1},
\end{aligned}
\end{equation}
where $d$ represents the inter-element spacing between two FPAs, and $\lambda$ denotes the carrier wavelength.

The SINR of communication user $k$ is calculated as
\begin{equation}\label{sinrk}
\operatorname{SINR}_k(\bW)=\frac{|\mathbf h_k^H\mathbf w_{k}|^2}{\sum_{j\neq k}|\mathbf h_k^H\mathbf w_{j}|^2+\sigma_k^2}.
\end{equation}

\vspace{-2mm}
\subsection{Sensing Model}
The received echo signal at the BS is expressed as
\begin{equation}
\begin{aligned}
\by_0&=\alpha_0 \mathbf{a}_r(\theta_0,\phi_0,{\mathbf r}) \mathbf{a}_t^{T}(\theta_0,\phi_0)\bx+\bc+\bn_0\\
&=\alpha_0\bA(\theta_0,\phi_0,\br)\bx+\bc+\bn_0,
\end{aligned}
\end{equation}
where $\alpha_0\in\mathbb C$ is the channel coefficient influenced by the target radar cross section and the path loss, $\theta_0$ and $\phi_0$ are the elevation and azimuth angles of the radar target, $\mathbf{a}_r(\theta_0,\phi_0,{\mathbf r})$ is the receive steering vector which will be discussed in detail later, $\bc\sim\mathcal{CN}(0,\bR_c)$ is the clutter with covariance matrix $\bR_c$ \cite{9724174}, and $\bn_0\sim\mathcal{CN}(0,\sigma_0^2\bI)$ is the AWGN at the BS. 

We model the clutter $\bc$ to be signal-dependent, given by
\begin{equation}
\bc=\sum_{i=1}^I\alpha_i\bA(\theta_i,\phi_i,\br)\bx,
\end{equation} 
where $I$ represents the number of clutters, $\theta_i$ and $\phi_i$ denote the elevation and azimuth angles of the $i$-th clutter, and $\alpha_i$ is the channel coefficient of the $i$-th clutter. Therefore, the covariance matrix of $\bc$ is given by 
\begin{equation}
\begin{aligned}
\bR_c=\sum_{i=1}^I|\alpha_i|^2\bA(\theta_i,\phi_i,\br)\bW\bW^H\bA(\theta_i,\phi_i,\br)^H.
\end{aligned}
\end{equation} 

For receiving of the BS, given the elevation and azimuth angles of arrival (AoAs) $\theta_0$ and $\phi_0$, the propagation distance difference between $\br_n$ and origin $\mathbf r_0=(0,0)^T$ is \cite{10243545,10318061}
\begin{equation}
\rho_0(\mathbf{r}_n)=x_n \sin \theta_0 \cos \phi_0+y_n \cos \theta_0.
\end{equation}
The signal phase difference between $\br_n$ and $\br_0$ is obtained as $2\pi\rho_0(\mathbf{r}_n)/\lambda$. By exploiting the far-field model, the receive steering vector is written as
\begin{equation}
\mathbf{a}_r(\theta_0,\phi_0,{\mathbf r})=\left[e^{\jmath\frac{2\pi}{\lambda}\rho_{0}(\mathbf{r}_1)},\dots,e^{\jmath\frac{2\pi}{\lambda}\rho_{0}(\mathbf{r}_N)}\right]^T \in \mathbb{C}^{N\times 1}.
\end{equation}

It is assumed that a linear filter $\bv$ is applied at the BS for maximizing the output SCNR \cite{9537599}. Then the output signal is expressed as 
\begin{equation}
\begin{aligned}
y_r=\bv^H\by_0=\alpha_0\bv^H\bA(\theta_0,\phi_0,\br)\bx+\bv^H\bc+\bv^H\bn_0.\label{radarre}
\end{aligned}
\end{equation}
The optimal $\bv$ is given by \cite{9724174}
\begin{equation}
\begin{aligned}
\bv=\beta_0(\bR_c+\sigma_0^2\bI)^{-1}\bA(\theta_0,\phi_0,\br)\bx,
\end{aligned}
\end{equation}
where $\beta_0$ denotes an arbitrary constant. By substituting $\bv$ into \eqref{radarre}, the radar SCNR can be expressed as \cite{9724174}
\begin{equation}
\begin{aligned}
\operatorname{SCNR}&=\mathbb{E}\left\{\frac{|\alpha_0|^2|\bv^H\bA(\theta_0,\phi_0,\br)\bx|^2}{\bv^H\bR_c\bv+\sigma_0^2\bv^H\bv}\right\}\\
&=|\alpha_0|^2\operatorname{tr}\left(\bA(\theta_0,\phi_0,\br)^H\bJ^{-1}\bA(\theta_0,\phi_0,\br)\bW\bW^H\right),\\
\end{aligned}
\end{equation}
where
\begin{equation}
\bJ=\sum_{i=1}^I|\alpha_i|^2\bA(\theta_i,\phi_i,\br)\bW\bW^H\bA(\theta_i,\phi_i,\br)^H+\sigma_0^2\bI.
\end{equation}

\subsection{Problem Formulation}
Since the radar SCNR dominates the detection probability, we choose it to evaluate the sensing performance \cite{10364735}. This paper focuses on maximizing the radar SCNR while adhering to constraints on the communication SINR and fluid antenna positions. The optimization problem is written as
\begin{equation}
\begin{aligned}
\max _{\bW, \br}
&\quad\operatorname{SCNR} \\
{\rm s.t.}
& \quad\operatorname{tr}(\bW\bW^H) \leq P_0,\\
& \quad \frac{|\mathbf h_k^H\mathbf w_{k}|^2}{\sum_{j\neq k}|\mathbf h_k^H\mathbf w_{j}|^2+\sigma_k^2}\geq \gamma_k, \quad\forall k,\\
&\quad\left\|\br_n-\br_l\right\|_2 \geq D, \quad 1\leq n \neq l \leq N,\\
& \quad{\mathbf r} \in \mathcal{S},\label{ori_p}
\end{aligned}
\end{equation}
where $P_0$ is the power budget, $\gamma_k$ represents the required SINR for user $k$, and $D$ specifies the minimum distance between adjacent fluid antennas to avoid mutual coupling effects.

Unfortunately, it is challenging to handle \eqref{ori_p} because it has coupled optimizing variables, non-convex constraints, and a non-concave objective function with inverse operation.

Evidently, the performance of FAS is highly dependent on the accuracy of CSI, and effective CSI acquisition methods are important for unleashing the capabilities of FAS. Recent efforts have seen several channel estimation schemes developed for FAS. For example, \cite{10236898} and \cite{10375559} applied compressed sensing methods to estimate and reconstruct CSI for FAS-assisted point-to-point and multiuser systems, respectively.

In this paper, we investigate the maximization problem of the radar sensing SCNR in the FAS-assisted MIMO ISAC system which has not been considered before. As a major first step, we assume perfect CSI at the BS. We note that some existing works have addressed system performance evaluation and optimization when instantaneous CSI is unavailable. For example, \cite{10328751} focused on maximizing the achievable rate of a FAS-assisted point-to-point system using statistical CSI. However, the impact of imperfect CSI on the system performance is out of the scope of this study. Future work could explore the effects of CSI errors on the system SCNR.

\vspace{-2mm}
\section{Proposed Algorithm}
In this section, we propose an iterative algorithm to find a locally optimal solution of the SCNR maximization problem. Specifically, the variables $\bW$ and $\br$ are first decoupled using the alternating optimization (AO) approach. After that, the transmit beamforming and receive antenna position sub-problems are addressed using the majorization-minimization (MM) method.

\vspace{-2mm}
\subsection{Transmit Beamforming Matrix}
Given the receive fluid antennas positions $\mathbf r$, the optimization sub-problem becomes
\begin{equation}
\begin{aligned}
\max _{\bW} 
&\quad\operatorname{SCNR} \\
{\rm s.t.} 
& \quad\operatorname{tr}(\bW\bW^H) \leq P_0,\\
& \quad \frac{|\mathbf h_k^H\mathbf w_{k}|^2}{\sum_{j\neq k}|\mathbf h_k^H\mathbf w_{j}|^2+\sigma_k^2}\geq \gamma_k, \quad\forall k.\\\label{tra}
\end{aligned}
\end{equation}

The inverse operation in the objective function SCNR is difficult to handle. Hence, we resort to the MM algorithm to find a lower bound as the surrogate function, given by \cite{10319318}
\begin{multline}\label{lowerw}
\operatorname{SCNR}\geq 2 \operatorname{Re}\left(\operatorname{tr}\left(\bW_p^H\bA^H \mathbf{J}_p^{-1} \bA\bW\right)\right)\\
-\operatorname{tr}\left(\mathbf{J}_p^{-1} \bA\bW_p \bW_p^H\bA^H\mathbf{J}_p^{-1} \mathbf{J}\right), 
\end{multline}
where $p$ is the iteration index for the MM algorithm. By using the lower bound, the objective function becomes concave. Then we can handle the non-convex SINR constraints. After expanding the constraints, they can be rewritten as
\begin{equation}
(1+\gamma_k^{-1})\mathbf h_k^H\mathbf w_{k}\mathbf w_{k}^H\mathbf h_k\geq \sum_j\mathbf h_k^H\mathbf w_{j}\mathbf w_{j}^H\mathbf h_k+\sigma_k^2, ~~\forall k.
\end{equation}
By employing the MM method to the left side of the inequality, the SINR constraints can be transformed to
\begin{multline}
(1+\gamma_k^{-1})\left(2\operatorname{Re}\left(\mathbf h_k^H\mathbf w_{k,p}\mathbf w_{k}^H\mathbf h_k\right)-\mathbf h_k^H\mathbf w_{k,p}\mathbf w_{k,p}^H\mathbf h_k\right)\\
\geq\mathbf h_k^H\bW\bW^H\mathbf h_k+\sigma_k^2,~~\forall k.
\end{multline}
Then the problem in \eqref{tra} can be reformulated as
\begin{equation}
\begin{aligned}
\max _{\bW} 
&\quad 2 \operatorname{Re}\left(\operatorname{tr}\left(\bW_p^H\bA^H \mathbf{J}_p^{-1} \bA\bW\right)\right) 
 \\&\quad-\operatorname{tr}\left(\mathbf{J}_p^{-1} \bA\bW_t \bW_p^H\bA^H\mathbf{J}_p^{-1} \mathbf{J}\right) \\
{\rm s.t.} 
& \quad\operatorname{tr}(\bW\bW^H) \leq P_0,\\
& \quad(1+\gamma_k^{-1})\left(2\operatorname{Re}\left(\mathbf h_k^H\mathbf w_{k,p}\mathbf w_{k}^H\mathbf h_k\right)-\mathbf h_k^H\mathbf w_{k,p}\mathbf w_{k,p}^H\mathbf h_k\right)\\&\quad\geq \mathbf h_k^H\bW\bW^H\mathbf h_k+\sigma_k^2, \quad\forall k,\label{finalW}
\end{aligned}
\end{equation}
which is a convex problem that can be tackled using conventional convex optimization methods.

\vspace{-2mm}
\subsection{Receive Fluid Antenna Position}
Given the transmit beamforming matrix $\mathbf W$, the original SCNR maximization problem can be recast into
\begin{equation}
\begin{aligned}
\max _{\br} 
&\quad\operatorname{SCNR} \\
{\rm s.t.} 
&\quad\left\|\br_n-\br_l\right\|_2 \geq D, \quad 1\leq n \neq l \leq N,\\
& \quad{\mathbf r} \in \mathcal{S}.
\end{aligned}
\end{equation}
Similarly, to tackle the inverse operation in SCNR, we obtain the lower bound by exploiting the MM algorithm, i.e.,
\begin{multline}\label{lowerr}
\operatorname{SCNR}\geq 2 \operatorname{Re}\left(\operatorname{tr}\left(\bW^H\bA_v^H \mathbf{J}_v^{-1} \bA\bW\right)\right)\\
-\operatorname{tr}\left(\mathbf{J}_v^{-1} \bA_v\bW \bW^H\bA_v^H\mathbf{J}_v^{-1} \mathbf{J}\right), 
\end{multline}
where $v$ is the iteration index for the MM algorithm. Then to deal with the coupled variable $\br$, we optimize one variable $\br_n$ while keeping others $\{\br_l,l\neq n\}_{l=1}^N$ fixed. We first expand the expression of lower bound in \eqref{lowerr} through vector multiplications, which is on the bottom of the next page.
\begin{figure*}[hb]
\centering\hrulefill
\begin{multline}\label{expand}
2 \operatorname{Re}\left(\mathbf{b} \mathbf{a}_r(\theta_0, \phi_0, \mathbf{r})\right)-\operatorname{tr}\left(\bE\left[\sum_{i=1}^I p_i|\alpha_i|^2\mathbf{a}_r(\theta_i,\phi_i,{\mathbf r})\mathbf{a}_r^H(\theta_i,\phi_i,{\mathbf r})\right]\right) \\
= 2 \operatorname{Re}\left(\sum_{n=1}^N b_n e^{\jmath\frac{2\pi}{\lambda}\rho_{0}(\mathbf{r}_n)} \right)-\sum_{i=1}^I p_i|\alpha_i|^2\sum_{n=1}^N \sum_{l=1}^Ne^{-\jmath\frac{2\pi}{\lambda}\rho_{i}(\mathbf{r}_l)}\bE_{ln} e^{\jmath\frac{2\pi}{\lambda}\rho_{i}(\mathbf{r}_n)}
\end{multline}
\end{figure*}
With fixed $\{\br_l,l\neq n\}_{l=1}^N$, maximizing the lower bound in \eqref{expand} is the same as maximizing ${q}\left(\br_n\right)$ on the bottom of the next
page, where
\begin{figure*}[hb]
\centering\hrulefill
\begin{multline}\label{q}
{q}\left(\br_n\right)=2|\bb_n|\cos\left(\angle \bb_n+\frac{2\pi}{\lambda}\rho_{0}(\br_n)\right)-
\sum_{i=1}^I|\alpha_i|^2 p_i\left(\bE_{nn}+2\sum_{l\neq n}^N|\bE_{ln}|\cos\left(\angle\bE_{ln}+\frac{2\pi}{\lambda}\left(\rho_{i}(\br_n)-\rho_{i}(\br_l)\right)\right)\right)
\end{multline}
\end{figure*}
\begin{align}
\bb&=\ba_t^T(\theta_0,\phi_0)\bW\bW^H\bA_v^H\bJ_v^{-1},\\
p_i&=\ba_t^T(\theta_i,\phi_i)\bW\bW^H\ba_t^{*}(\theta_i,\phi_i),\\
\bE&=\mathbf{J}_v^{-1} \bA_v\bW \bW^H\bA_v^H\mathbf{J}_v^{-1}.
\end{align}

Since ${q}\left(\br_n\right)$ is a non-concave function, the MM method is exploited to derive a lower bound, given by \cite{10243545}
\begin{equation}
\begin{aligned}
{q}\left(\br_n\right) &\geq {q}\left(\br_{n,c}\right)+\nabla {q}\left(\br_{n,c}\right)^T\left(\br_n-\br_{n,c}\right)\\&-\frac{\delta_n}{2}\left(\br_n-\br_{n,c}\right)^T\left(\br_n-\br_{n,c}\right) \\
&=-\frac{\delta_n}{2} \br_n^T \br_n+\left(\nabla{q}\left(\br_{n,c}\right)+\delta_n \br_{n,c}\right)^T \br_n+{q}\left(\br_{n,c}\right)\\&-\frac{\delta_n}{2}\left(\br_{n,c}\right)^T \br_{n,c},\label{rllow}
\end{aligned}
\end{equation}
where $c$ is the iteration index for the inner MM algorithm, $\delta_n$ is a positive real number satisfying $\delta_n\bI_2\succeq\nabla^2{q}(\br_n)$, $\nabla {q}(\br_n)$ and $\nabla^2{q}(\br_n)$ represent the gradient vector and Hessian matrix of ${q}(\br_n)$ with respect to $\br_n$, respectively. Additionally, the expression of $\delta_n$ is written as \cite{10243545}
\begin{equation}
\begin{aligned}
&\delta_n=\frac{16\pi^2}{\lambda^2}\left(|\bb_n|+\sum_{i=1}^I |\alpha_i|^2 p_i\sum_{l\neq n}^N|\bE_{ln}|\right).\label{deltan}
\end{aligned}
\end{equation}

Now, according to the lower bound in \eqref{rllow}, we need to maximize $\tilde{q}\left(\br_n\right)=-\frac{\delta_n}{2} \br_n^T \br_n+\left(\nabla {q}\left(\br_{n,c}\right)+\delta_n \br_{n,c}\right)^T \br_n$. Therefore, the corresponding optimization problem related to $\br_n$ can be written as
\begin{subequations}\label{rsub}
\begin{align}
\max _{\mathbf{r}_n} 
&\quad -\frac{\delta_n}{2} \br_n^T \br_n+\left(\nabla {q}\left(\br_{n,c}\right)+\delta_n \br_{n,c}\right)^T \br_n\label{q2_1}\\
{\rm s.t.} 
&\quad{\mathbf{r}_n} \in \mathcal{S}, \label{cons_2_1} \\
&\quad \left\|\mathbf{r}_n-\mathbf{r}_l\right\|_2 \geq D, \quad 1\leq l \neq n \leq N.\label{cons_2_2} 
\end{align}
\end{subequations}
\begin{algorithm}[t]
\color{black}
\caption{MM-Based SCNR Maximization Algorithm}\label{a_1}
\begin{algorithmic}[1]
\State Initialize $\mathbf{W}$, $\br$, threshold $\varepsilon$, set iteration index $t=0$, $p=0$, $v=0$, $c=0$.
\Repeat 
\Repeat
\State Obtain $\bW_{p+1}$ by optimizing \eqref{finalW}.
 \State Set $p=p+1$.
\Until The increase of SCNR is below a threshold $\varepsilon$.
\Repeat 
\For {$n = 1: N$}
\Repeat
\State Calculate $\delta_n$ as \eqref{deltan}.
\State Update $\br_{n,c+1}$ according to \eqref{ropt1}.
\If {$\br_{n,c+1}$ does not satisfy \eqref{cons_2_1} or \eqref{cons_2_2}}
\State Update $\br_{n,c+1}$ by optimizing \eqref{finalr}
\EndIf
\State Set $c=c+1$.
\Until The increase of \eqref{lowerr} is below a threshold $\varepsilon$.
\EndFor
\State Set $v=v+1$.
\Until The increase of SCNR is below a threshold $\varepsilon$.
 \State Set $t=t+1$.
 \Until The increase of SCNR is below a threshold $\varepsilon$.
 \end{algorithmic}
\end{algorithm}
If we ignore the constraints \eqref{cons_2_1} and \eqref{cons_2_2}, then we can obtain the closed-form solution of $\br_n$, given by 
\begin{equation}\label{ropt1}
\br_{n,c+1}^{\mathrm{opt}}=\frac{1}{\delta_n}\nabla {q}\left(\br_{n,c}\right)+\br_{n,c}.
\end{equation}
Then we substitute \eqref{ropt1} into \eqref{cons_2_1} and \eqref{cons_2_2} to see whether the constraints are satisfied. If the constraints are satisfied, then $\br_{n,c+1}^{*}$ in \eqref{ropt1} is the optimal solution of \eqref{rsub}. If not, then we need to address the non-convex constraints \eqref{cons_2_2}. By applying the MM method, we can obtain the inequality given by \cite{10243545}
\begin{equation}
\begin{aligned}
&\left\|\mathbf{r}_n-\mathbf{r}_l\right\|_2  \geq
\frac{1}{\left\|\mathbf{r}_{n,c}-\mathbf{r}_l\right\|_2}\left(\mathbf{r}_{n,c}-\mathbf{r}_l\right)^T\left(\mathbf{r}_n-\mathbf{r}_l\right) .
\end{aligned}
\end{equation}
Then the sub-problem related to $\br_n$ can be reformulated as
\begin{subequations}\label{finalr}
\begin{align}
\max _{\mathbf{r}_n}
&\quad -\frac{\delta_n}{2} \br_n^T \br_n+\left(\nabla {q}\left(\br_{n,c}\right)+\delta_n \br_{n,c}\right)^T \br_n\\
{\rm s.t.}
&\quad{\mathbf{r}_n} \in \mathcal{S},  \\
&\quad \frac{1}{\left\|\mathbf{r}_{n,c}-\mathbf{r}_l\right\|_2}\left(\mathbf{r}_{n,c}-\mathbf{r}_l\right)^T\left(\mathbf{r}_n-\mathbf{r}_l\right) \geq D,\nonumber \\&\quad\quad\quad\quad\quad\quad\quad\quad\quad l=1,2, \ldots, N, \quad l \neq n.
\end{align}
\end{subequations}
The problem in \eqref{finalr} can be solved using conventional convex optimization methods because it is a convex problem.

\subsection{Convergence and Complexity Analysis}
\textbf{Algorithm \ref{a_1}} provides a summary of the proposed MM-based SCNR maximization algorithm. As illustrated in Algorithm 1, the variables $\mathbf{W}$ and $\mathbf{r}$ are alternatingly optimized. Specifically, $\mathbf{W}$ is updated using the MM method, and the sequence of objective values $\{\operatorname{SCNR}^{p}\}_{p=0}^{\infty}$ is guaranteed to converge according to \cite{7547360}. Thus, the proposed approach for optimizing $\mathbf{W}$ ensures that the system SCNR value does not decrease at each iteration of \textbf{Algorithm \ref{a_1}}. Furthermore, to make the problem related to the vector $\mathbf{r}$ more tractable, we have employed a combination of the MM and AO methods. In this approach, we iteratively optimize the position of one antenna at a time while keeping the others fixed, until convergence is reached. After obtaining the optimized positions for all antennas using this algorithm, the practical application involves moving all antennas simultaneously to their corresponding optimized positions. The MM-based method for optimizing $\mathbf{r}_n$ is also guaranteed to converge according to \cite{7547360}. Consequently, the proposed approach for optimizing $\mathbf{r}_n$ ensures non-decreasing values of $q(\mathbf{r}_n)$. As the objective function in (21) is non-decreasing, the AO algorithm for optimizing $\mathbf{r}$ is guaranteed to converge \cite{bezdek2003convergence}. Based on these considerations, both the solutions for $\mathbf{W}$ and $\mathbf{r}$ ensure that the SCNR objective value does not decrease, i.e., $\operatorname{SCNR}(\mathbf{W}^{t},\mathbf{r}^{t}) \geq \operatorname{SCNR}(\mathbf{W}^{t-1},\mathbf{r}^{t-1})$, where $t$ denotes the iteration index of the AO algorithm. Therefore, the overall convergence of the proposed methodology for jointly optimizing $\mathbf{W}$ and $\mathbf{r}$, as described in \textbf{Algorithm \ref{a_1}}, is guaranteed.

Similar to \cite{10319318}, updating $\bW$ has a computational complexity approximated as
\begin{multline}
\mathcal{O}\left((2K)^{1/2}MK(M^2K^2+KM^2+2M^2K^2) \right)\\
=\mathcal{O}(K^{3.5}M^3). 
\end{multline}
Moreover, updating $\br$ has a computational complexity that is given by 
\begin{equation}
\mathcal{O}\left(\left(N^4\zeta_1+M^{2.5}N \ln (1 / \mu) \zeta_2\right)\zeta_3\right), 
\end{equation}
where the variables $\zeta_1$, $\zeta_2$, and $\zeta_3$ represent the maximum number of iterations associated with Steps 10--15, Step 13, and Steps 8--18, respectively, and $\mu$ indicates the accuracy of the interior-point technique \cite{10243545}.  Consequently, we can determine the total computational complexity of \textbf{Algorithm \ref{a_1}} as 
\begin{equation}
\mathcal{O}\left(\left(K^{3.5}M^3+\left(N^4\zeta_1+M^{2.5}N \ln (1 / \mu) \zeta_2\right)\zeta_3\right)\zeta\right), 
\end{equation}
where $\zeta$ is the maximum number of outer iterations.

\section{Simulations Results}
The numerical results for evaluating the proposed algorithm are shown in this section. The MIMO ISAC system includes a BS with $M=8\times 8$ transmit antennas and $N=4$ receive fluid antennas, $K=4$ single-FPA communication users, a single radar target, and $I=9$ radar clutters. The carrier wavelength is set as $\lambda=0.015$ m. It is assumed that the elevation and azimuth AoAs/AoDs and radar target/clutter are randomly distributed within $[0,\pi/2]$. We assume there are $L_k=L=20$ paths on the communication side. The path gain $\rho_{k,l}$, the radar target channel coefficient $\alpha_0$, and the clutter channel coefficient $\alpha_i$ are modeled as a standard Gaussian distributed random variable \cite{9724174}. Also, $\gamma_k=\gamma=1$ is the needed SINR for user $k$. The noise variance $\sigma_k^2$ and $\sigma_0^2$ are set to be $-105$ dBm \cite{you2020energy}. We set $D=\lambda/2$ as the minimum distance \cite{10243545}. Moreover, the convergence threshold is set as $\varepsilon=10^{-4}$.

To benchmark the proposed FAS-assisted MIMO ISAC scheme, we consider the following baseline schemes:
\begin{itemize}
\item \textbf{Alternating position selection (APS)}:
The given square region $\mathcal S$ is discretized into positions with intervals of $D=\lambda/2$. An exhaustive search is performed within the discretized regions to find the optimal position.

\item \textbf{Rotatable uniform linear array (RULA)}: The receive antenna is a RULA with antenna spacing of $D=\lambda/2$. The rotation angle of the RULA is quantized into $50$ discrete angles, and an exhaustive search method is used to select the rotation angle that maximizes the SCNR.

\item \textbf{FPA}: The receive antennas are arranged in a planar grid with equal spacing $D=\lambda/2$ both horizontally and vertically between adjacent elements.
\end{itemize}

\begin{figure}[htbp]
\centering
\includegraphics[scale=0.49]{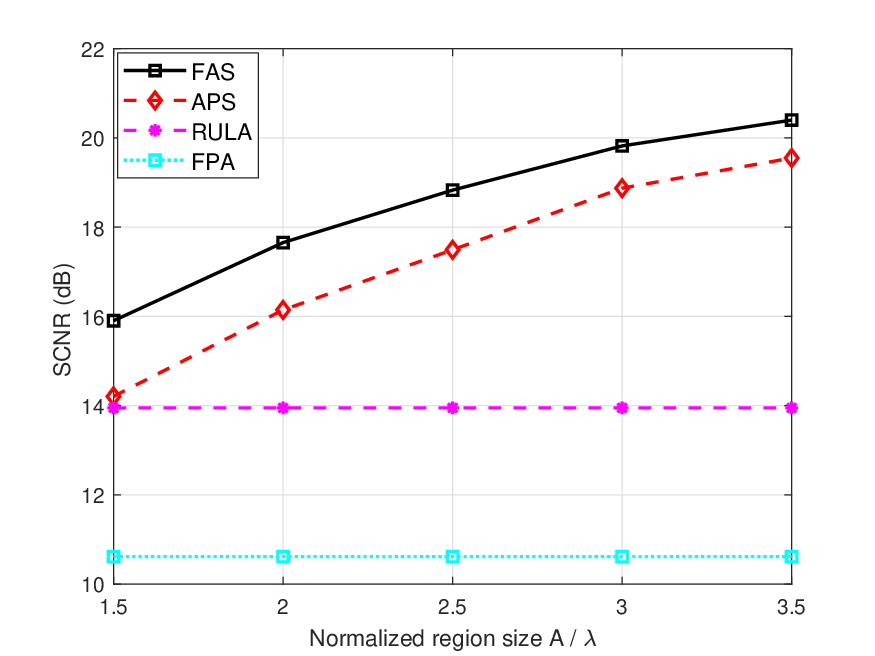}
\captionsetup{font=footnotesize, singlelinecheck=off}
\caption{SCNR versus the size of the normalized region.}\label{fig0}
\vspace{-2mm}
\end{figure}

\begin{figure}[htbp]
\centering
\includegraphics[scale=0.49]{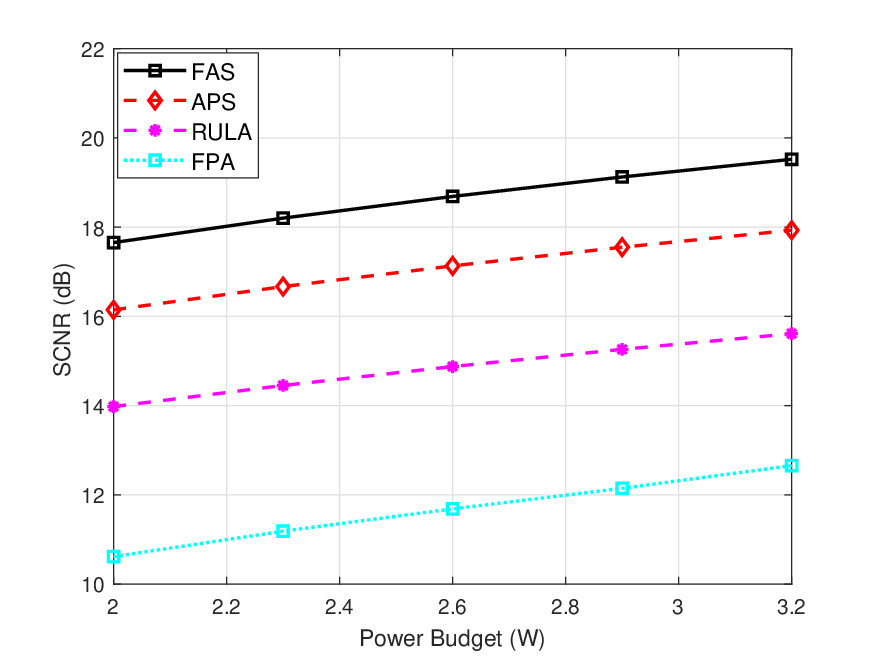}
\captionsetup{font=footnotesize, singlelinecheck=off}
\caption{SCNR versus the power budget.}\label{fig1}
\vspace{-2mm}
\end{figure}

\begin{figure}[!t]
\centering
\includegraphics[scale=0.49]{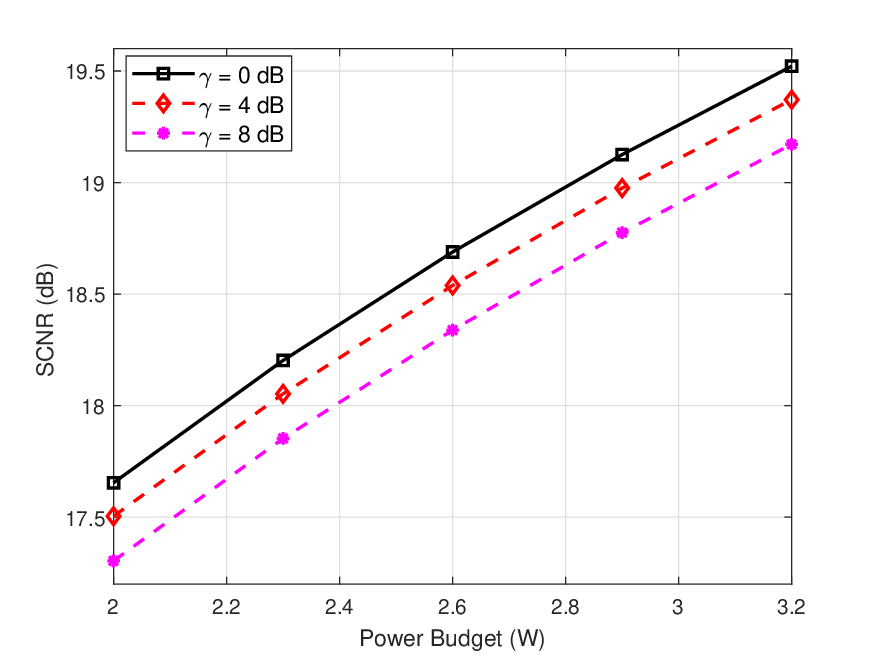}
\captionsetup{font=footnotesize, singlelinecheck=off}
\caption{SCNR versus the power budget for different $\gamma$.}\label{fig2}
\vspace{-2mm}
\end{figure}

In Fig.~\ref{fig0}, the results are presented to illustrate the SCNR performance of the FAS design and other baselines with regard to the size of the normalized region $A/\lambda$. It is evident that the FAS design consistently outperforms the other three baselines across all normalized region sizes. The reason for the superior performance of the proposed FAS design is its ability to alter the position of the antenna elements dynamically. Moreover, the FAS design shows a steady increase in SCNR as the size of the normalized region grows. The APS design demonstrates a gain in SCNR compared to RULA and FPA designs since the antennas can change among discrete positions.

Fig.~\ref{fig1} shows the relationship between SCNR and the power budget for the five designs. All of the designs exhibit an upward trend in SCNR as the power budget increases. This is expected as higher power budgets typically allow for better signal processing capabilities, leading to improved SCNR. The FAS design consistently exhibits the highest SCNR values across the entire range of power budgets. Results also show that APS and RULA provide moderate gains, whereas FPA consistently underperforms compared to the other designs.

Fig.~\ref{fig2} illustrates the relationship between power budget and SCNR under varying $\gamma$ parameters. As $\gamma$ increases, a noticeable reduction in the system's SCNR can be observed. This phenomenon occurs because, with higher $\gamma$, a larger portion of the available power is allocated to satisfy the SINR requirements for communication, leaving less power available for target sensing. This behavior highlights the tradeoff between communication SINR and radar SCNR, demonstrating how power allocation decisions impact both communication and sensing performance in the system.

\vspace{-2mm}
\section{Conclusion}\label{sec_conclusion}
In this paper, we investigated the joint beamforming and antenna position design in the FAS-assisted MIMO ISAC. In particular, the radar SCNR maximization under communication SINR and antenna position constraints was formulated. We first adopted the AO approach to decouple the variables in order to address the non-convex problem. Then the non-convex sub-problems were converted to convex ones using the MM method. Numerical results demonstrated that the FAS-assisted ISAC scheme attains superior SCNR performance.
\vspace{-2mm}

\bibliographystyle{IEEEtran}
%\bibliography{EE_AI}

\end{document}